\documentclass[11pt]{article}
\usepackage{deauthor,times}
\usepackage{graphicx}
\usepackage{comment}

\title{5G: Agent for Further Digital Disruptive Transformations}
\author{
\makebox[.3\linewidth]{Beng Chin Ooi} \\
\textit{ooibc@comp.nus.edu.sg}\\
National University of Singapore
\and 
\makebox[.2\linewidth]{Gang Chen} \\
\textit{cg@zju.edu.cn}\\
Zhejiang University
\and
\makebox[.3\linewidth]{Dumitrel Loghin}\\
\textit{dumitrel@comp.nus.edu.sg}\\
National University of Singapore 
\and
\makebox[.45\linewidth]{Wei Wang}\\
\textit{wangwei@comp.nus.edu.sg}\\
National University of Singapore
\and 
\makebox[.45\linewidth]{Meihui Zhang}\\
\textit{meihui\_zhang@bit.edu.cn}\\
Beijing Institute of Technoloy
}

\begin{document}

\maketitle

% Other opinion papers do not have abstract. I moved it to the Introduction.

\begin{comment}
\begin{abstract}

With 5G on the verge of being adopted as the next mobile network, it is timely
to analyze its impact on the landscape of computing, in particular on data
management and data-driven technologies. With a predicted increase of
10-100$\times$ in bandwidth and 5-10$\times$ decrease in latency, 5G is expected
to be the main enabler for edge computing which includes accessing cloud-like
services, as well as conducting machine learning at the edge. We analyze its
impact on both traditional and emerging technologies, and discuss future
research challenges and opportunities.

\end{abstract}
\end{comment}

\begin{abstract}

The fifth-generation (5G) mobile communication technologies are on the way to be
adopted as the next standard for mobile networking. It is therefore timely to
analyze the impact of 5G on the landscape of computing, in particular, data
management and data-driven technologies. With a predicted increase of
10-100$\times$ in bandwidth and 5-10$\times$ decrease in latency, 5G is expected
to be the main enabler for edge computing which includes accessing cloud-like
services, as well as conducting machine learning at the edge. In this paper, we
examine the impact of 5G on both traditional and emerging technologies, and
discuss research challenges and opportunities.

\end{abstract}

\vspace{-3mm}
\section{Introduction}
\vspace{-1mm}

% Intro

% Overview of 5G
5G specifications are handled by the 3rd Generation Partnership Project (3GPP),
while the actual implementation is done by big networking hardware players, such
as Nokia, Ericsson, Huawei, Qualcomm, among others. Compared to the current 4G
technologies which are widely-spread all over the world, 5G is supposed to have
a higher bandwidth of up to 10 Gbps, lower latency of 1 ms and a higher device
density of up to one million devices per square
kilometer~\cite{5g_our_article,5g_ch_opp}. 5G operates in a high-frequency band
between 28 GHz and 95 GHz, also known as the millimeter wave spectrum
(mmWave)~\cite{5g_our_article,5g_ch_opp}. While this spectrum allows for larger
bandwidths, 5G also employs massive multiple-input and multiple-output
(MIMO)~\cite{5g_our_article} technology to further increase the bandwidth. MIMO
uses large antenna arrays in both the base station and the device to allow for
parallel data streams and to direct the radio wave such that it avoids
interference and achieves superior spectral efficiency~\cite{5g_our_article}.
Consequently, 5G is supposed to be more energy-efficient compared to current
wireless technologies.

% 5G technologies
5G does not bring only improved communication speeds, but also a series of
technologies that have the potential to change the computing landscape in a
disruptive way. Among these technologies, we distinguish Software Defined
Networking (SDN), Network Function Virtualization (NFV), Network Slicing (NS),
and Device-to-Device communications (D2D)~\cite{edge_mec_survey_2017}. SDN
represents methods to separate the data plane, which is responsible for handling
and forwarding networking packets, and the control plane, which is responsible
for establishing the route of the packets. NFV represents the usage of commodity
hardware running virtualized services to replace custom networking hardware. For
example, a commodity server could run firewall services instead of using a
specialized physical firewall. Network Slicing enables several logical networks
to share a single physical network infrastructure. D2D communication is a
feature of 5G that allows devices to communicate directly, with minimum help
from a central authority. For example, the base station may help only with
device pairing and authentication, while subsequent steps, including data
transfers, are performed without its involvement. In this paper, we group SDN,
NFV, and NS into 5G virtualization, while D2D is a distinct feature.

\vspace{-1mm}
\section{Digital Disruptive Transformations}
\label{sec:impact}

\begin{figure}[tp]
\centering
\includegraphics[width=\textwidth]{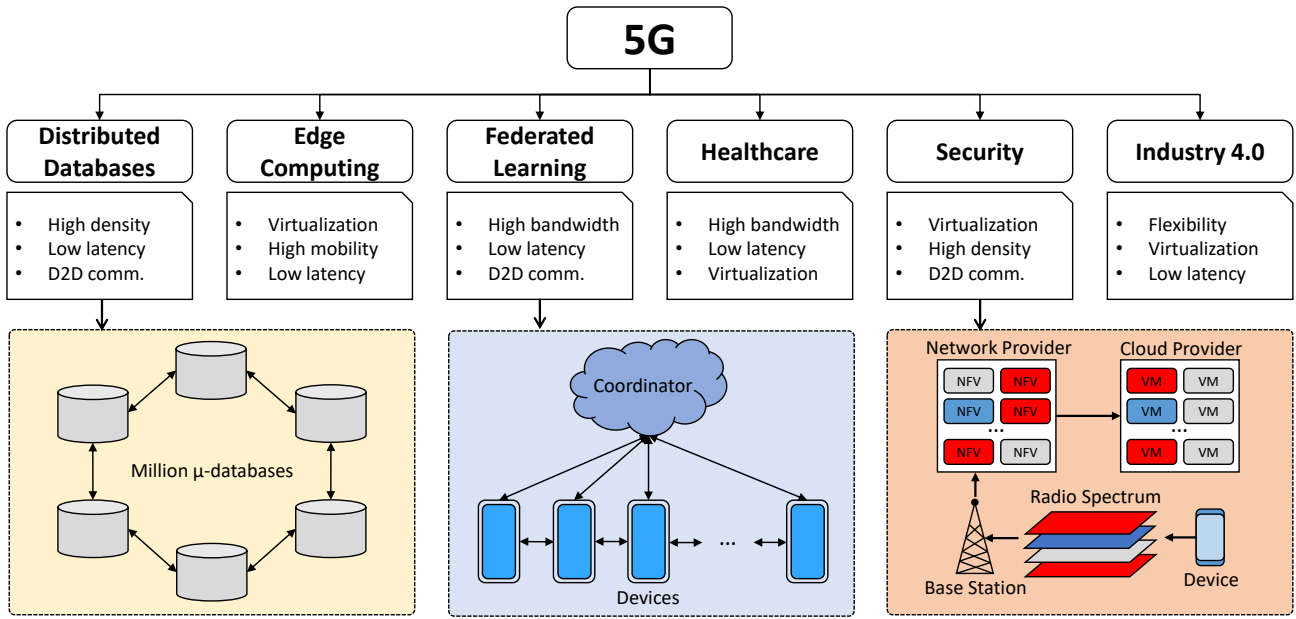}
%\includegraphics[bb = 0 0 888 421,
% width=\textwidth]{figs/5GImpact_Bulletin.png}
\caption{The impact of 5G on different domains. Some 5G features have higher impact on each domain.}
\label{fig:5gimpact}
\end{figure}

\vspace{-1mm}
Among different domains that are going to be significantly impacted by the
adoption of 5G~\cite{5g_our_article}, we discuss three key areas related to data
management and data-driven technologies, as highlighted in
Figure~\ref{fig:5gimpact}.

\vspace{-2mm}
\subsection{Distributed and Federated Data Processing}
\vspace{-1mm}

With the increasing number of data breaches and awareness of the General Data
Protection Regulation (GDPR) and value of data, the demand for having full
control of the data by the user is on the rise. For example, the healthcare
records of a patient may be stored in the individual's mobile device instead of
being fragmented and stored only in hospital databases. 5G has the potential to
bring to reality the concept of millions of micro-databases with each being kept
in an individual edge device, in the form of distributed and federated
micro-databases, as shown in Figure~\ref{fig:5gimpact}. GDPR and federated data
processing in dynamic networks due to node churning and joining introduce many
new challenges such as accuracy, completeness, fairness, and representativeness,
in addition to performance issues.

The key characteristics of 5G that could help with the implementation of
federated micro-databases are high device density, low latency, energy
efficiency, and D2D communications. For instance, a 5G deployment is expected to
be able to support one million devices in a square kilometer~\cite{5g_ch_opp}.
With a few millions of interconnected devices, a smart city becomes the
playground of federated micro-databases. Low latency and D2D communication allow
nodes to communicate fast and directly, in a setup that may prove to be superior
even to datacenters with Gigabit Ethernet links. Compared to these datacenters,
a distributed 5G setup has a few advantages, such as improved networking, higher
energy efficiency, and mobility.

First, D2D communication without direct base station involvement is reducing the
risk of network partitioning due to faulty centralized infrastructure. In
comparison, a datacenter that depends on a few switches and routers is more
prone to partitioning. Second, 5G terminals are predicted to be more
energy-efficient~\cite{5g_ch_opp}. This, together with the low-power nature of
smartphones and IoT devices, could help in reducing the energy consumption by up
to 10$\times$ compared to a classic datacenter based on high-performance
servers~\cite{5g_our_article}.

In the context of increased enthusiasm for blockchain technologies, we analyze
the impact of 5G in this domain which is closely-connected with distributed
databases~\cite{anh_blockchain_tkde}. A blockchain ledger represents a database
distributed across thousands or millions of physical nodes, in a Byzantine
environment where peers do not trust each other since some of them may be
malicious. Currently, mobile networks are not involved in blockchains because
the nodes are most likely connected via wired or optical links. At most, some
clients interacting with blockchain peers may use mobile devices. But with the
increasing scalability issue of the blockchain, and the adoption of solutions
involving sharding~\cite{hung_sharding} or second-tier,
sub-blockchains~\cite{lightning_bitcoin}, 5G has the potential to impact the
performance of these systems~\cite{5g_our_article}.
Shards or second-tier blockchains may run at the edge of the network and include
both fixed and mobile nodes and clients and peers may also run on the same
physical node at the edge.

\vspace{-2mm}
\subsection{Federated Learning}
\vspace{-1mm}

The explosion of mobile and IoT devices at the edge requires a new approach
towards efficient machine learning (ML). These devices act as both data
consumers (e.g. actuators) and data producers (e.g. sensors). As data consumers,
these devices run model inference on their own collected data. As data
producers, these devices push data to higher network levels, where more powerful
systems run ML model training~\cite{singa15}. But the explosion of edge devices
exerts too much pressure on the networking connections to the cloud, and on
cloud's computation and storage resources~\cite{5g_our_article}. A solution to
this problem is \textit{federated learning}.

Federated learning~\cite{federated_learning_arxiv_16} entails the building of a
model on multiple devices that contribute their data to the training process. A
coordinator gets the learned parameters from the devices to build an aggregated
model, as shown in Figure~\ref{fig:5gimpact}. This approach is directly
addressing the issue of isolated data islands, where data is found in different
locations, under different organizations, and it cannot be merged or aggregated.

We envision that 5G is going to accelerate the adoption of federated learning.
With high bandwidth and low latency, local model parameters and the aggregated
model can be shared much faster between the devices and the coordinator. D2D
communication could relieve some pressure from the device-coordinator
connections by sharing intermediate parameters directly. However, this D2D
communication introduces security risks in environments with malicious devices.
On the other hand, network virtualization could help in solving the security and
privacy issues by creating isolated slices for the task of federated learning.
\vspace{-2mm}
\subsection{Security}
\vspace{-1mm}

% DDoS
The adoption of 5G is going to create new security challenges. We analyze these
challenges based on the 5G characteristics involved. First, we discuss the
higher device density, higher bandwidth and lower latency that could create the
ideal environment for launching massive distributed denial of service (DDoS)
attacks~\cite{ddos_botnet}. It is well known that IoT devices are relatively
easier to compromise compared to servers, due to factors such as low system
performance that does not allow running complex anti-virus solutions on the
device, software immaturity and bad security practices which are adopted to get
faster time-to-market. With 5G allowing more IoT devices to be connected to the
Internet, the attack surface is going to increase significantly. One of the
biggest attacks to date was done using infected IoT devices with a botnet called
Mirai~\cite{ddos_botnet} which targeted Dyn DNS servers and took down many
websites, especially on the East side of the USA.

% D2D comms -> authentication -> blockchain
Secondly, we examine the impact of D2D communications on security. D2D is
supposed to reduce the traffic to base stations, but will require strict
security protocols to avoid privacy violations and device hijacking. For
example, D2D communications may require an ad-hoc authentication step to
determine the identity of the devices. Given the scale of 5G networks, a
centralized solution is unfeasible. We envision an authentication service based
on the decentralized blockchain to avoid data tempering. However, current
blockchains suffer from low throughput and high latency, hence there is a need
for developing novel blockchain platforms.

% network slicing
Thirdly, we analyze the impact of network slicing on security. As a
generalization of virtualization, network slicing allows different applications
to share the same physical network by operating across all layers of the
networking stack. At the physical layer, the radio connection is multiplexed
through spectrum sharing. At the networking layer, providers use SDN and NFV to
multiplex the network. At the application level, computing resources are
multiplexed using virtual machines (VM), either on the cloud or at the edge. This multitude
of virtualized resources managed by different parties is a challenge for
security. The threats could be present at all layers, as shown in
Figure~\ref{fig:5gimpact} where the honest user (blue) is attacked by malicious
actors (red). Achieving the isolation of the entire slice across all layers
poses a significant challenge because there is a need to apply a cross-layer
coordinated security protocol.

\vspace{-2mm}
\section{Conclusions}
\label{sec:conclusions}
\vspace{-1mm}

In summary, the adoption of 5G is expected to accelerate the development of
emerging technologies, such as IoT, edge computing, blockchain, and federated
learning. In addition, 5G is going to give rise to new systems, such as millions
of interconnected databases, and generate new use cases, such as remote work,
immersive augmented reality, telemedicine and smart automotive, among
others~\cite{5g_our_article}.
Security is one of the key challenges of end-to-end virtualization in 5G
networks. It remains to be studied how to ensure security across systems managed
by different entities and threatened by different security risks. Another key
challenge is ensuring data privacy in the context of millions of interconnected
databases and federated learning.
\vspace{1mm}
\\
\noindent \textbf{Acknowledgement}: This research is supported by Singapore
Ministry of Education Academic Research Fund Tier 3 under MOE’s official grant number
MOE2017-T3-1-007.

\vspace{-8pt}

% un-comment for dev
\begin{comment}
\small
\bibliographystyle{ieeetr}
\bibliography{references}
\end{comment}

\end{document}